\documentclass{article}

\usepackage{arxiv}

\usepackage[utf8]{inputenc} 
\usepackage[T1]{fontenc}    
\usepackage{hyperref}       
\usepackage{url}            
\usepackage{booktabs}       
\usepackage{amsfonts}       
\usepackage{nicefrac}       
\usepackage{microtype}      
\usepackage{lipsum}
\usepackage{soul}
\usepackage{color}
\usepackage{float}
\usepackage{graphicx}
\usepackage{caption}
\usepackage{subcaption}
\usepackage{titlesec}
\usepackage{multirow}
\usepackage{amsmath}
\usepackage{soul,color}
\usepackage[section]{placeins}
\usepackage{array, makecell} %
\usepackage{float}
\usepackage[section]{placeins}
\usepackage{wrapfig}
\usepackage{enumitem}

\usepackage[nodisplayskipstretch]{setspace}

\title{Crystal structure prediction of materials with high symmetry using differential evolution}
\author{
   \And
 Wenhui Yang\\
 School of Mechanical Engineering\\
  Guizhou University \\
  Guiyang China 550025 \\
  \And
    Edirisuriya M. Dilanga Siriwardane, Rongzhi Dong\\
  Department of Computer Science and Engineering\\
  University of South Carolina\\
  Columbia, SC 29201 \\
  \And
  Yuxin Li\\
  School of Mechanical Engineering\\
  Guizhou University\\
  Guiyang China 550025 \\    \And
 Jianjun Hu \thanks{Corresponding author: J.H. (http://www.cse.sc.edu/~jianjunh)}\\
  Department of Computer Science and Engineering\\
  University of South Carolina\\
  Columbia, SC, 29201, USA \\
  \texttt{jianjunh@cse.sc.edu}
}

\begin{document}
\maketitle

\begin{abstract}

Crystal structure determines many properties of materials. With the crystal structure of a chemical substance, many physical and chemical properties can be predicted by first-principles calculations or machine learning models. Since it is relatively easy to generate a hypothetical chemically valid formula, crystal structure prediction becomes an important method for discovering new materials. In our previous work, we proposed a contact map-based crystal structure prediction method, which uses global optimization algorithms such as genetic algorithms to maximize the match between the contact map of the predicted structure and the contact map of the real crystal structure to search for the coordinates at the Wyckoff Positions(WP), demonstrating that known geometric constraints (such as the contact map of the crystal structure) help the crystal structure reconstruction. However, when predicting the crystal structure with high symmetry, we found that the global optimization algorithm has difficulty to find an effective combination of WPs that satisfies the chemical formula, which is mainly caused by the inconsistency between the dimensionality of the contact map of the predicted crystal structure and the dimensionality of the contact map of the target crystal structure. This makes it challenging to predict the crystal structures of high-symmetry crystals. In order to solve this problem, here we propose to use PyXtal to generate and filter random crystal structures with given symmetry constraints based on the information such as chemical formulas and space groups. With contact map as the optimization goal, we use differential evolution algorithms to search for non-special coordinates at the Wyckoff positions to realize the structure prediction of high-symmetry crystal materials. Our experimental results show that our proposed algorithm CMCrystalHS can effectively solve the problem of inconsistent contact map dimensions and predict the crystal structures with high symmetry. 
\end{abstract}

\keywords{crystal structure prediction \and random crystal structure \and contact map \and differential evolution \and high symmetry }

\section{Introduction}
Discovering novel materials with special properties has great impact over new product design such as high-capacity lithium batteries, superconductors, and solar panel materials with high efficiency. With the advancement of materials genome research, big data, and artificial intelligence, new materials have promoted the development of high-tech\cite{kvashnin2019computational,he2020computational,brown2020artificial} and emerging industries. To speed up the discovery of new materials, materials informatics has emerged as a new material discovery model that uses data-driven methods\cite{gomez2018automatic,jablonka2020big,hoffmann2019data} to build machine learning models for large-scale screening and generation of materials. There are several approaches for discovering new materials including first-principles\cite{curtis2018gator,woodley2008crystal,maddox1988crystals} calculation based structural tinkering, inverse materials design\cite{zunger2018inverse,kim2020inverse}, generative machine learning models\cite{kim2020inverse,dan2020generative,bradshaw2019model,noh2019inverse,ren2020inverse, dan2020generative} and computational crystal structure predictions\cite{glass2006uspex,kvashnin2019computational,ryan2018crystal,hu2021contact}.

The structure of a material determines its many properties. In the past, the way to obtain materials' structures mostly relied on experimental X-ray diffraction(XRD)\cite{oviedo2019fast}. However, the huge composition space of the materials  makes it too inefficient to use these methods to design and explore new materials. Materials scientists have made breakthroughs in crystal structure prediction methods based on a combination of quantum mechanics and structure search algorithms \cite{oganov2011evolutionary}, which has accelerated the design of new materials. However,  the structure search space is still too huge for the structure prediction of complex compounds and existing crystal structure prediction methods are still inefficient. While global search algorithms combined with first-principle free energy calculations have discovered a series of new materials\cite{oganov2011evolutionary,wang2020calypso}, they usually rely on expensive DFT calculations of the free energies of the sampled structures\cite{zhang2017materials,oganov2019structure} and can only be used for structure prediction of relatively small systems. To improve the performance of crystal structure prediction, various strategies such as clustering, intelligent mutation operators, and active learning\cite{podryabinkin2019accelerating,lookman2019active} have been proposed to improve sampling efficiency.

Recently, we have developed a generative machine learning model (MatGAN) based on generative adversarial networks (GAN)\cite{dan2020generative} which can generate new hypothetical inorganic materials formulas by learning the implicit chemical composition rules to form compounds. However, it is important to obtain their structures to evaluate their physicochemical properties. To do that, we propose a new machine learning based crystal structure prediction framework \cite{hu2021contact,hu2020distance}, which first predicts the contact map or distance map of hypothetical materials and then use these maps to reconstruct the coordinates of the unit cell atoms. Compared with the crystal structure prediction methods based on global free energies optimization\cite{oganov2011evolutionary,lyakhov2013new}, our methods take advantage of a large number of learned, hidden constraints, components and atomic configuration rules and constraints in the known crystal structures, which can make the sampling of the search space more effective.

In our previous work\cite{0Contact} on contact map-based crystal structure prediction, we use global optimization algorithms such as GA and PSO to maximize the match between the contact map of the predicted structure and the contact map of the real crystal structure to search for the coordinates at the Wyckoff positions in the unit cell, proving that known geometric constraints (such as the contact map of the crystal structures) can help the crystal structure reconstruction. However, when predicting the crystal structure of materials with high structural symmetry (e.g.Fcc crystal structures), we found that the global optimization algorithm usually faces the difficulty of dealing with the inconsistent dimensions of the contact map of the predicted crystal structure and the contact map of the target crystal structure due to the degenerate overlapping atomic positions after symmetric operations. The problem of inconsistent dimensions makes it challenging to predict high symmetric crystal structures. The crystal structures with high symmetry usually contain many special fractional coordinates (0,1/6, 1/4, 1/3, 1/2, 2/3, 3/4, 5/6, -3/8, -1/4, -1/8,etc.), and the multiplicity of these special positions is smaller than the multiplicity of general positions. The global optimization algorithm mainly searches for general positions, and it is difficult to search for these special positions as they require the precise special fractional coordinates. This will cause the WP combination searched by the global optimization algorithm to be incompatible with the chemical formula of the target crystal material, and the dimensions of the contact map are thus inconsistent, making it difficult to calculate the difference/error between the predicted contact map and the given target contact map. In our previous crystal structure prediction based on contact map, we can only predict the crystal structures of low symmetry, requiring that the multiplicity of WP is equal to the number of symmetry transformation operations and the structure does not contain positions with special fractional coordinates that may lead to degenerated positions. The resulting structure reconstruction algorithm thus has very limited coverage. In order to solve this problem, here we propose to use PyXtal\cite{fredericks2021pyxtal}(a Python software package that helps crystal structure generation and crystal symmetry analysis) to generate random crystal structures with given symmetry constraints based on information such as chemical formula and space group. In order not to destroy the symmetry of the random crystal structure, we fixed the special coordinate values (0, 1/4, 1/3, 1/2, 2/3, 3/4, etc.), and only optimize the non-special coordinates in the Wyckoff positions using the differential evolution algorithm. Our new algorithm is named CMCrystalHS (Contact map based structure prediction of crystals with high symmetry).

Our contribution in this paper can be summarized as follows:

\begin{itemize}

    \item We identify the challenge of contact map based crystal structure prediction for highly symmetric structures due to the inconsistent dimensionality of the contact map caused by the degenerated atomic positions
    \item We propose a novel symmetric template generation and differential evolution algorithm based optimization to predict the crystal structures of high symmetry
    \item We evaluate our algorithms on different types of symmetric structures with extensive experiments to demonstrate the effectiveness of our algorithm for reconstructing the crystal structures of high symmetry from the contact map. 
    \item We apply our algorithms to predict the crystal structures for a set of hypothetical formulas and discovered several new stable materials structures as verified by DFT formation energy calculations.
\end{itemize}

\section{Methods}

\subsection{Contact map based crystal structure prediction}

 Crystal structure information can be specified by its lattice constant, space group and element types and coordinates at the Wyckoff positions. Lattice constants are important parameters of crystalline materials, which are closely related to the binding energy between atoms. Changes in the lattice constants will affect the internal composition and force state of the crystal. Space group is a collection of all symmetry elements in the internal structure of a crystal. The space group type and symmetry type reflect the symmetry of the atoms in the internal crystal structure. There are 230 space groups among all crystal structures. The Wyckoff positions are used to indicate the symmetry of the equivalent atoms in the unit cell. The contact map records the connection information between all atoms in the crystal structure and captures the interaction between atoms. 
 
 In our previous work\cite{0Contact}, we developed a contact map based crystal structure prediction framework in which the space group, the lattice constants, and contact map of atoms in the unit cells can all be predicted using data driven machine learning models \cite{zhao2020machine,li2020mlatticeabc,0Contact,hu2021alphacrystal}. With these information, we have demonstrated that it is possible to reconstruct the coordinates of the Wyckoff positions using a global optimization algorithm \cite{0Contact}. The optimization algorithm reconstructs the crystal structure by searching the Wyckoff positions by maximizing the match between the contact map of the predicted structure and the contact map of the target crystal structure.

\begin{figure}[H] 
    \begin{subfigure}[t]{0.3\textwidth}
        \includegraphics[width=\textwidth]{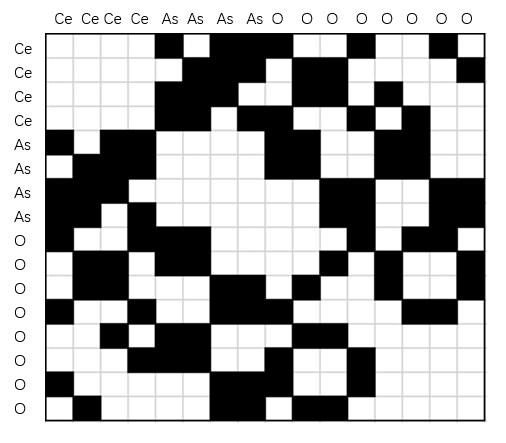}
        \caption{Ce2As2O6 (predicted contact map 16×16)}
        \vspace{-3pt}
        \label{fig:Ce2As2O6_predim}
    \end{subfigure}\hfill
    \begin{subfigure}[t]{0.3\textwidth}
        \includegraphics[width=\textwidth]{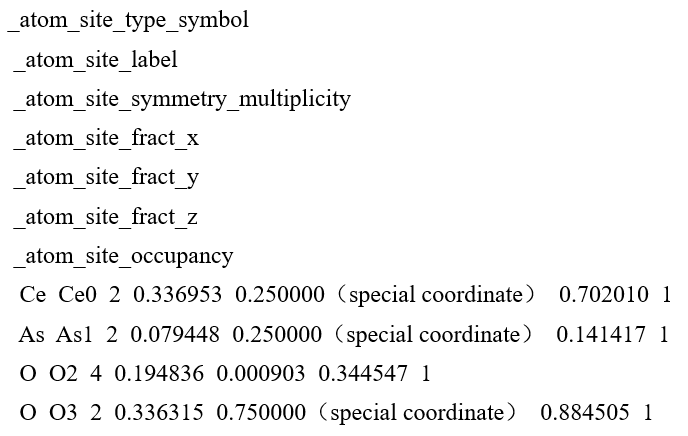}
        \caption{Ce2As2O6 (Wyckoff positions)}
        \vspace{-3pt}
        \label{fig:special_coordinates}
    \end{subfigure}\hfill
    \begin{subfigure}[t]{0.3\textwidth}
        \includegraphics[width=\textwidth]{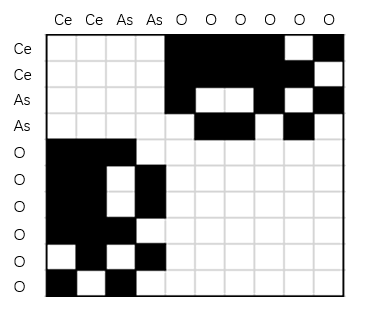}
        \caption{Ce2As2O6 (real contact map 10×10)}
        \vspace{-3pt}
        \label{fig:Ce2As2O6_trudim}
    \end{subfigure}
    \caption{The dimensions of the predicted contact map and the target contact map are inconsistent due to the special coordinates of Wackoff positions of highly symmetric crystal structures. }
    \label{fig:dim}
\end{figure}

For crystal structures with high symmetry, however, it is difficult for the optimization algorithm to find an effective combination of WPs that satisfy the chemical formula and have the same dimension of the target contact map, which leads to the failure of the structure prediction. As is shown in Figure\ref{fig:dim},for Ce2As2O6 whose space group number is 11, the dimension of the contact map of the predicted crystal structure is 16×16 for a candidate structure while the dimension of the real/final contact map is 10×10. The reason is that the target structure has special coordinates with high symmetry (multiple symmetry operations over the WP may map to the same atomic positions) while the structure under optimization has non-special coordinates which will be mapped to different atom positions after symmetric operations. The mismatched contact map dimensions will make it difficult to calculate meaningful distance between two contact maps.

\subsection{Prediction of crystal structures of high symmetry based on contact map}

\begin{figure}[ht]
  \centering
  \includegraphics[width=0.6\linewidth]{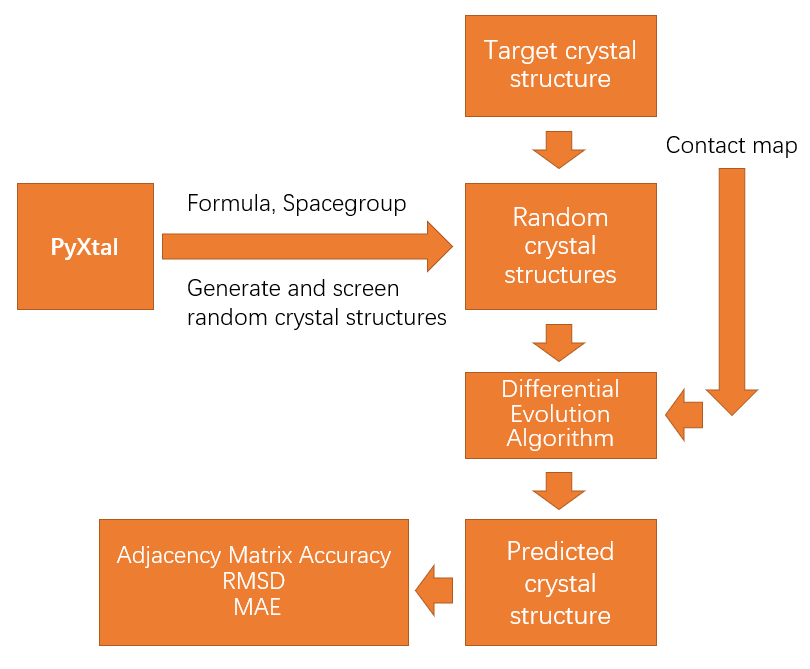}
  \caption{Prediction framework of high-symmetry crystal structures based on contact map.}
  \label{fig:framework}
\end{figure}

Here we propose a novel approach for reconstructing crystal structures of materials with high symmetry. Our prediction framework is shown in Figure\ref{fig:framework}. First, we use the PyXtal library to generate 50 random crystal structures of high symmetry based on the material formula and space group. Next, a suitable random crystal structure of high symmetry is selected based on its distance to the target contact map. Finally, the selected template structure will be optimized by the  the differential evolution algorithm to search for appropriate non-special coordinates that can minimize the difference of the predicted contact map and the target contact map. The prediction performance will be evaluated using the accuracy of the contact map, the root mean square distance (RMSD) and the mean absolute error (MAE) between the predicted Wyckoff positions of the crystal structure and the Wyckoff positions of the target structure to evaluate the reconstructed crystal structure.

\subsection{Generating random crystal structures}

PyXtal \cite{fredericks2021pyxtal} is a Python software package that helps design the material structure with a certain symmetry constraint, which is useful for crystal structure prediction and crystal symmetry analysis. It can generate atomic structures for a given symmetry and stoichiometry and do geometric optimization. The PyXtal library uses the space group and its Wyckoff positions as a template, inserting atoms (or molecules) at the Wyckoff position one at a time until the given stoichiometry is met, thereby generating random symmetrical crystals. In this way, the correct symmetry can be obtained without adjusting the positions of the atoms. During this process, it also restricts the atoms from getting too close. The following are the steps for PyXtal to generate a random crystal,as is shown in Figure\ref{fig:pyxtalstructure}:

\begin{wrapfigure}{R}{6cm}
    \centering
    \includegraphics[width=0.9\linewidth]{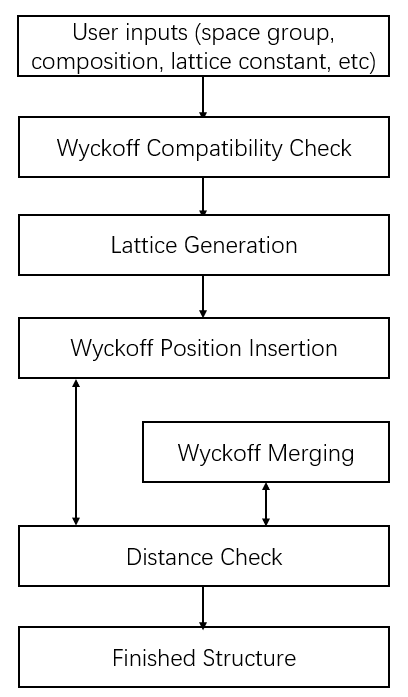}
    \vspace{-1em}
    \caption{Random structure generation process by PyXtal given the stoichiometry and symmetry}
    \vspace{-1em}
    \label{fig:pyxtalstructure}
\end{wrapfigure}

\begin{enumerate}[label=(\Roman*)]
  \item Get user input parameters, such as chemical formula, space group, etc. The user can also define the lattice and the allowable distance between atoms (suitable for high pressure). 
  \item Compatibility check of stoichiometry and space group. For different space groups, the Wyckoff positions have different multiplicities. Therefore, some atomic numbers may be incompatible with the space group. PyXtal searches for all possible Wyckoff position combinations within the stoichiometric range to find a valid combination from them and fails to try to generate a random crystal structure if it cannot be found. Some space groups allow effective combinations of Wyckoff positions, but may result in a lack of many degrees of freedom when generating random crystal structures, resulting in atoms being very close. Nevertheless, PyXtal will also try to generate before the maximum limit is reached or the generation is successful. 
  \item Generate a random lattice corresponding to the space group. The random crystal structure generation of PyXtal starts from the selection of the unit cell. The symmetry group matches a specific type of lattice. To avoid ambiguity, all crystals use conventional unit cell selection. Triclinic cells are the most common case, and other cell types can be obtained by using various constraints.
  \item Start by placing atoms in the Wyckoff positions, inserting one atomic specie at a time. The crystal structure in PyXtal is mainly composed of Wyckoff positions (WP). After determining the space group and lattice, you can grow the structure by inserting one Wyckoff position at a time. PyXtal starts with the largest available WP (general position of the space group). When the required number of atoms is equal to or greater than the size of the regular position, the algorithm continues, otherwise the next largest Wyckoff position (or Wyckoff position set) is selected in descending order to ensure that the larger position is selected as far as possible, and the greater the multiplicity, the higher the prevalence rate. 
  \item If the minimum distance of an atom to other atoms at a single Wyckoff position is less than a certain tolerance (based on the atomic species), the Wyckoff position is merged into a smaller special position. In order to make the resulting structure have realistic bonds and bond lengths, the distance between the generated atoms should not be too close. PyXtal defaults that the distance between two atoms should not be shorter than the length of the covalent bond between them. The user can also define the allowable distance between any two types of atoms, and decide the appropriate shorter or longer cut-off distance. 
  \item If the newly generated Wyckoff position is too close to the previously generated position, a new generation point is selected within the same original Wyckoff position.
  \item Continue to insert Wyckoff positions and species until the correct stoichiometry is met. If any step in the above steps III-VI fails, the operation will be repeated before the maximum number of attempts is reached. 
\end{enumerate}

In our algorithm, after a random crystal structure is successfully generated, we will fix the special coordinates of corresponding WP positions and then try to search the non-special coordinates to maximize the match of its contact map with the target contact map.

\subsection{3D crystal structure reconstruction algorithm}

Contact map based crystal structure prediction problem can be mapped as a global optimization problem. Here we select Nevergrad\cite{nevergrad}, which is an open source platform for derivative-free optimization. It contains various optimizers (such as NGOpt, DE, PSO, etc.), and supports multi-objective optimization and handles constraints. Here, we choose the differential evolution (DE) optimizer to search for the Wyckoff positions guided by a given contact map to perform the prediction of its corresponding crystal structure.

Differential Evolution algorithms (DE)\cite{price2006differential} are a population-based adaptive global optimization algorithms, which are simple, efficient, and converge quickly. The basic idea of the algorithm is: first, generate a random initial population, and sum the vector difference of any two individuals in the population with the third individual to generate a new individual. If the fitness of the new individual is better than the current individual, replace the old individual with the new individual in the next generation, otherwise the old individual will remain. The optimal solution is approached by continuous iteration.

For all the DE based experiments, we set the population size to 100 and the number of generations to 1000 with a cross-over probability of 0.5 and F1=0.8, F2=0.8.
\subsection{Evaluation metrics}

The contact map M of the crystal structure is converted from the distance matrix in the pymatgen library according to the distance threshold between atoms in the VESTA software, for each pair of atoms in the unit cell, if their distance is within the range of $[Min.length,Max.length]$, then there is a bond between them and the corresponding contact map position M[i,j] is set to 1, if not it is set to 0. The objective function of the differential evolution algorithm is as follows: 

\begin{equation}
\operatorname{fitness}_{opt}=\frac{2|A \cap B|}{|A|+|B|} \approx\frac{2 \times A \bullet B}{\operatorname{Sum}(A)+\operatorname{Sum}(B)}
\end{equation}

where$A$ is the predicted contact map and $B$ is the true contact map of a given composition, both only contain 1/0 entries.  $A \cap B$  denotes the common elements between A and B, |g| represents the number of elements in a contact map, • denotes dot product, Sum(g) is the sum of all contact map elements. The fitness essentially measures the overlap of two contact map samples, with values ranging from 0 to 1 with 1 indicating perfect overlap. We call this performance measure as contact map accuracy too.

To evaluate the reconstruction performance of DE, we can use the contact map accuracy as one evaluation criterion. Additionally, we define the root mean square distance (RMSD) and mean absolute error (MAE) of two crystal structures as below:
\begin{equation}
    \begin{aligned}
\mathrm{RMSD}(\mathbf{v}, \mathbf{w}) &=\sqrt{\frac{1}{n} \sum_{i=1}^{n}\left\|v_{i}-w_{i}\right\|^{2}} \\
&=\sqrt{\frac{1}{n} \sum_{i=1}^{n}\left(\left(v_{i x}-w_{i x}\right)^{2}+\left(v_{i y}-w_{i y}\right)^{2}+\left(v_{i z}-w_{i z}\right)^{2}\right)}
\end{aligned}
\end{equation}

\begin{equation}
        \begin{aligned}
\mathrm{MAE}(\mathbf{v}, \mathbf{w}) &=\frac{1}{n} \sum_{i=1}^{n}\left\|v_{i}-w_{i}\right\| \\
&=\frac{1}{n} \sum_{i=1}^{n}\left(\|v_{i x}-w_{i x}\|+\|v_{i y}-w_{i y}\|+\|v_{i z}-w_{i z}\|\right)
\end{aligned}
\end{equation}

where $n$ is the number of independent atoms in the true crystal structure. For symmetrized structures, $n$ is the number of independent atoms of the set of Wyckoff equivalent positions. For regular structures, it is the total number of atoms in the structure. $v_i$ and $w_i$ are the corresponding atoms in the predicted crystal structure and the true crystal structure. 

\section{Experimental Results}
\label{sec:headings}

\subsection{Generating and screening random crystal structures with a given symmetry}

First we would like to show the quality of random crystal generation given the composition and space group. We selected a set of crystal structures from the Materials Project database as our test cases with the space group ranging from 42 to 221. We use PyXtal to generate 50 random crystal structures for each chemical formula and its space group of the target crystal material. PyXtal can generate a random crystal lattice consistent with the space group, but the error may be larger. We can set the crystal lattice to that of the target crystal structure when generating the random crystal structure. The purpose of our experiment is to show that known geometric constraints (such as the contact map of the crystal structure) can contribute to the reconstruction of the crystal structure. Since PyXtal can generate many types of random crystal structures, how to select a suitable crystal structure from the 50 random crystal structures as the template structure for further DE based coordinate optimization guided by the contact map is a key issue. During the experiment, we found that PyXtal can generate many different random crystal structures of WP combinations according to a chemical formula and its space group. The number and multiplicity of WPs in some random crystal structures are inconsistent with the target crystal structure, but the number of their total atoms is consistent. In these cases, the structures can also be optimized based on the contact map to improve the accuracy of the contact map, but the random crystal structure WP after optimization may have a larger error compared with the target crystal structure.

We use the following principles to select a suitable random crystal structure from the generated 50 random crystal structures for optimization:

\begin{itemize}
    \item The number of WPs is the least (that is, the multiplicity of each WP is as large as possible)
    \item The multiplicity of the WP of each atom is arranged in descending order
    \item The random crystal structure with the largest contact map accuracy is selected from the random crystal structures that meet the above two conditions
\end{itemize}

The WP combination of the random crystal structure screened by this method is likely to be consistent with the WP combination of the target crystal structure. 

\begin{figure}[H] 
    
    \begin{subfigure}[t]{0.4\textwidth}
        \includegraphics[width=\textwidth]{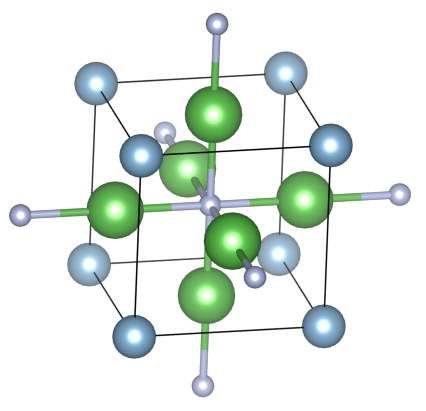}
        \caption{La3Al1N1 (random crystal structure)}
        \vspace{-3pt}
        \label{fig:La3Al1N1_random}
    \end{subfigure}\hfill
    \begin{subfigure}[t]{0.4\textwidth}
        \includegraphics[width=\textwidth]{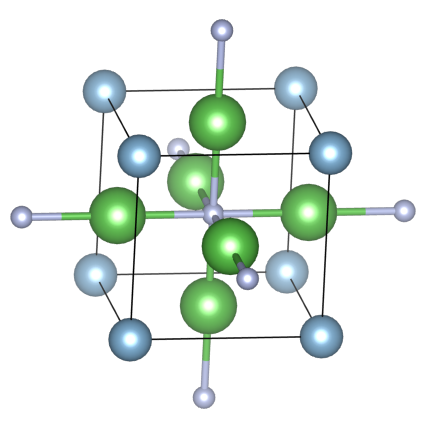}
        \caption{La3Al1N1 (real)}
        \vspace{-3pt}
        \label{fig:La3Al1N1}
    \end{subfigure}
    
    \caption{Comparison of generated random crystal structures with the given composition and space group and real crystal structures}
    \label{fig:random_real}
\end{figure}

Figure\ref{fig:random_real} shows material La\textsubscript{3}Al\textsubscript{1}N\textsubscript{1} with its real structure and corresponding random structure generated by PyXTal. The random structure have the contact map accuracy of 100\% selected out of 50 random crystal structures. The random crystal structure is consistent with the atom connections of the target crystal structure. The random crystal structure's WP combination of La\textsubscript{3}Al\textsubscript{1}N\textsubscript{1} is consistent with the target crystal structure, which shows that when WP error is small and it is easier to obtain high contact map accuracy.

\begin{table}[!htb] 
\begin{center}
\caption{Target crystal structures}
\label{table:structure_performance}
\begin{tabular}{|l|l|l|l|l|l|l|l|l|l|}
\hline
\textbf{\makecell{target}} & 
\textbf{\makecell{mp\_id}} & 
\textbf{\makecell{space\\group}} &
\multicolumn{1}{c|}{\textbf{no.of WPs}} &
\textbf{\makecell{atoms\\in unit\\cell}} &   \multicolumn{1}{c|}{\textbf{\makecell{no.of\\ special \\coordinates}}} & 
\multicolumn{1}{c|}{\textbf{\makecell{no.of\\ non-special \\coordinates}}}\\ \hline

Ce\textsubscript{4}Se\textsubscript{8}             & mp-1040452    & 42               & 2        & 12       & 3             & 3   \\ \hline
Tb\textsubscript{2}Cd\textsubscript{4}F\textsubscript{16}              & mp-29201    & 82               & 4        & 22     & 5    & 7         \\ \hline
Fe\textsubscript{4}O\textsubscript{4}F\textsubscript{4}              & mp-1178215    & 84               & 3        & 12      & 3             & 6   \\ \hline
Nb\textsubscript{4}Cr\textsubscript{4}O\textsubscript{16}              & mp-758053    & 109               & 4        & 24     & 6    & 6         \\ \hline

K\textsubscript{1}Ti\textsubscript{6}Se\textsubscript{8}              & mp-1223931    & 147               & 4        & 15     & 5    & 7         \\ \hline
Cr\textsubscript{3}N\textsubscript{6}              & mp-1096914    & 154               & 2        & 9     & 2    & 4         \\ \hline
In\textsubscript{2}I\textsubscript{6}O\textsubscript{18}              & mp-23400    & 173               & 5        & 26      & 2             & 13   \\ \hline
Ho\textsubscript{2}H\textsubscript{6}O\textsubscript{6}              & mp-24075    & 176               & 3        & 14     & 5    & 4         \\ \hline
Li\textsubscript{2}Cr\textsubscript{2}O\textsubscript{4}              & mp-753227    & 186               & 4        & 8     & 8    & 4         \\ \hline
La\textsubscript{3}Al\textsubscript{1}N\textsubscript{1}              & mp-4505    & 221               & 3        & 5     & 9    & 0         \\ \hline
\end{tabular}
\end{center}
\end{table}

\subsection{Prediction results of crystal structure based on contact map}

To evaluate the crystal structure reconstruction performance of our CMCrystalHS algorithm, we select 10 test cases from the Materials Project database as shown in Table 1. The space group numbers range from 42 to 221 and the number of Wyckoff positions are between 2 and 4. These structures also have different number atoms ranging from 5 to 26 and different number of non-special coordinates from 0 to 13. For all the DE optimizations, the running times range from 300 seconds to 1000 seconds, depending on the complexity of the structures (the number of non-special coordinates to optimize).

Table \ref{table:overall_performance_compare} shows the crystal structure reconstruction performance of our algorithm. For all the test targets, we report the predicted structure accuracy only using 50 random structure generation (before optimization) and after DE based coordinate optimization. First we find that by using only PyXtal based random structure generation, the contact map accuracy can reach as high as 100\% for La\textsubscript{3}Al\textsubscript{1}N\textsubscript{1}, which is a highly symmetric cubic $\operatorname{Pm} \overline{3} m$ structure (space group 221) and contains only special coordinates. For this target, no improvement is possible by our DE algorithm. For the remaining targets, the contact map accuracy range from 0.593 to 0.818 where the lowest accuracy is from In\textsubscript{2}I\textsubscript{6}O\textsubscript{18}, which has the largest number of non-special coordinates and the maximum number of atoms in the unit cell. Accordingly, the RMSD errors range from 0.0 to 0.309 and the MAE errors are between 0.0 and 0.183. 

In the middle\textit{Optimized} columns, we show the prediction accuracy after applying contact map guided DE based WP coordinate optimization. We find that for four targets(Ce\textsubscript{4}Se\textsubscript{8}, Fe\textsubscript{4}O\textsubscript{4}F\textsubscript{4} ,K\textsubscript{1}Ti\textsubscript{6}Se\textsubscript{8}, and Li\textsubscript{2}Cr\textsubscript{2}O\textsubscript{4}), we have increased their contact map accuracy to 100\% with 25\%, 49.93\%, 49.93\%, 33.33\% improvement respectively, which significantly demonstrates the effectiveness of contact map guided coordinate reconstruction for crystal structure prediction. The all other remaining 4 targets, the contact map accuracy improvements are between 6.67\% and 37\%. At the same time, the WP coordinates RMSD errors have also been reduced from 1.05\% to 39.67\%. In terms of MAE error, there is only one case that the DE optimization increases the error for  In\textsubscript{2}I\textsubscript{6}O\textsubscript{18} by 5.59\%. For all other cases, the MAE errors have been reduced by significantly ranging from 6.88\% to 40.94\%. 

\begin{figure}[hb!]
	\centering
	
    \begin{subfigure}{.29\textwidth}
		\includegraphics[width=\textwidth]{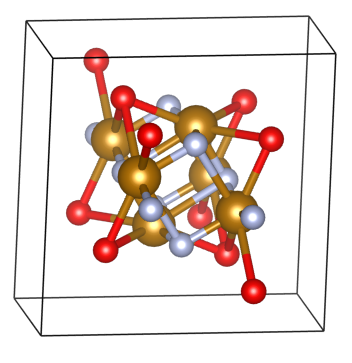}
		\caption{Random crystal structure of Fe\textsubscript{4}O\textsubscript{4}F\textsubscript{4} with contact map accuracy:66.7\%, RMSD: }
		\vspace{3pt}
	\end{subfigure}
    \begin{subfigure}{.32\textwidth}
		\includegraphics[width=\textwidth]{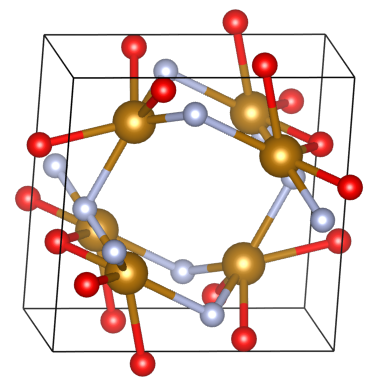}
		\caption{Predicted structure of Fe\textsubscript{4}O\textsubscript{4}F\textsubscript{4} with contact map accuracy:100\%, RMSD: }
		\vspace{3pt}
	\end{subfigure}
	\begin{subfigure}{.35\textwidth}
		\includegraphics[width=\textwidth]{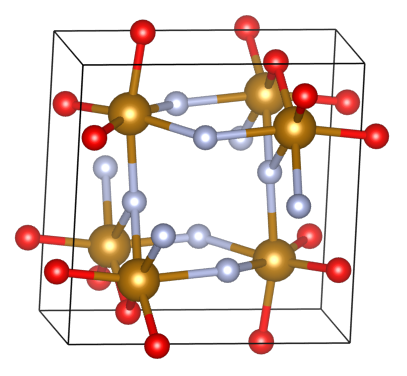}
		\caption{Target structure Fe\textsubscript{4}O\textsubscript{4}F\textsubscript{4}}
		\vspace{3pt}
	\end{subfigure}
	
	\begin{subfigure}{.32\textwidth}
		\includegraphics[width=\textwidth]{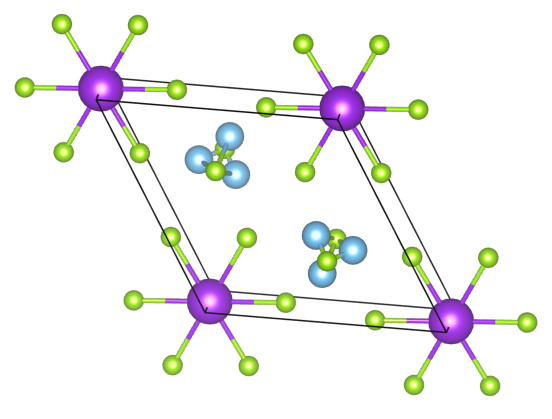}
		\caption{Random crystal structure of K\textsubscript{1}Ti\textsubscript{6}Se\textsubscript{8} with contact map accuracy:66.7\%, RMSD:0.117 }
		\vspace{3pt}
	\end{subfigure}
	\begin{subfigure}{.32\textwidth}
		\includegraphics[width=\textwidth]{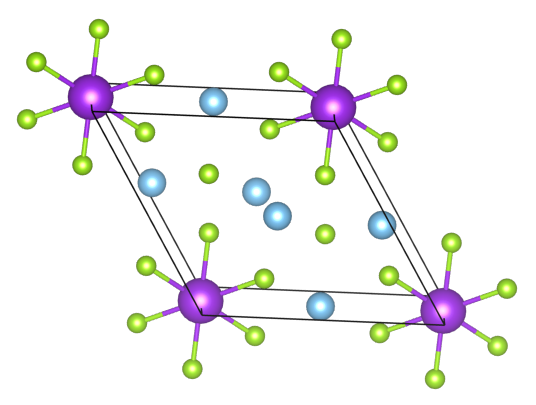}
		\caption{Predicted structure of K\textsubscript{1}Ti\textsubscript{6}Se\textsubscript{8} with contact map accuracy:100\%, RMSD:0.072 }
		\vspace{3pt}
	\end{subfigure}
	\begin{subfigure}{.32\textwidth}
		\includegraphics[width=\textwidth]{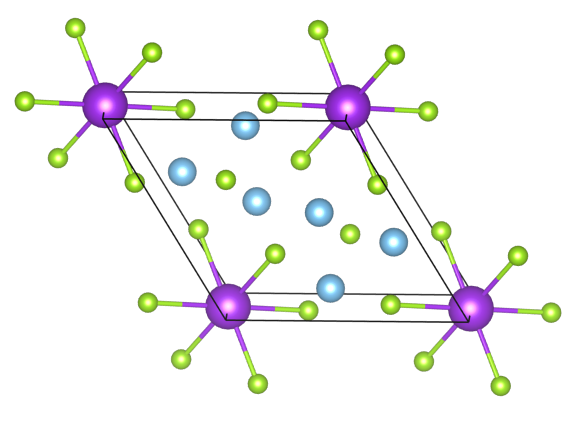}
		\caption{Target structure K\textsubscript{1}Ti\textsubscript{6}Se\textsubscript{8}}
		\vspace{3pt}
	\end{subfigure}
	
	\begin{subfigure}{.19\textwidth}
		\includegraphics[width=\textwidth]{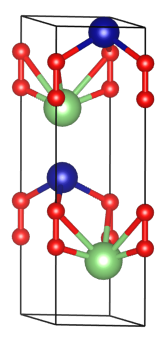}
		\caption{Random crystal structure of Li\textsubscript{2}Cr\textsubscript{2}O\textsubscript{4} with contact map accuracy:\%, RMSD: }
		\vspace{3pt}
	\end{subfigure}
	\begin{subfigure}{.19\textwidth}
		\includegraphics[width=\textwidth]{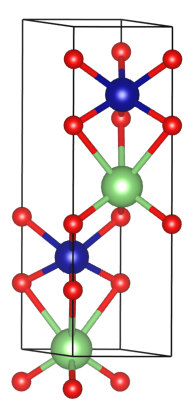}
		\caption{Predicted structure of Li\textsubscript{2}Cr\textsubscript{2}O\textsubscript{4} with contact map accuracy:100\%, RMSD: }
		\vspace{3pt}
	\end{subfigure}
	\begin{subfigure}{.2\textwidth}
		\includegraphics[width=\textwidth]{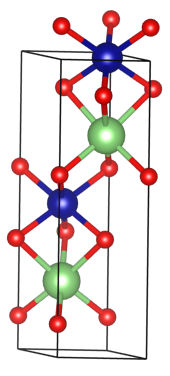}
		\caption{Target structure Li\textsubscript{2}Cr\textsubscript{2}O\textsubscript{4}}
		\vspace{3pt}
	\end{subfigure}
	
	\caption{Experimental results of crystal structure prediction based on contact map}
	\label{fig:predictedstructures}
\end{figure}

To further demonstrate how the DE-based optimization helps to reconstruct the crystal structures, we show the three case studies in Figure \ref{fig:predictedstructures}. Figure \ref{fig:predictedstructures} (a)-(c) shows the prediction results for Fe\textsubscript{4}O\textsubscript{4}F\textsubscript{4} with a space group 84. The symmetry-guided random structure generation creates a structure with only 66.7\% contact map accuracy compared to the true structure with many atoms packed together. After DF-based non-special coordinates optimization guided by the contact map, the contact map accuracy is increased to 100\% and its structure is very close to the true structure in (c). The DE-optimization also helps the random structure of K\textsubscript{1}Ti\textsubscript{6}Se\textsubscript{8} generated by PyXtal in Figure \ref{fig:predictedstructures}(d) to be fine-tuned into 100\% contact map accuracy by unpacking the clustered atoms in (d). More dramatic structural change is also possible to increase the contact map accuracy as shown in Figure \ref{fig:predictedstructures} (g) and (h).

\begin{table}[H] 
\begin{center}
\caption{Prediction performance comparison before and after DE based crystal structure optimization}
\label{table:overall_performance_compare}
\begin{tabular} {|l|l|l|l|l|l|l|l|l|l|}
\hline
                    & \multicolumn{3}{c|}{Before optimization}                                                      & \multicolumn{3}{c|}{Optimized}                                               & \multicolumn{3}{c|}{Improvement \%}                                                              \\ \hline
material           & \multicolumn{1}{c|}{\makecell{contact\\map\\ accuracy}} & \multicolumn{1}{c|}{RMSD} & \multicolumn{1}{c|}{MAE} & \multicolumn{1}{c|}{\makecell{contact\\map\\accuracy}} & \multicolumn{1}{c|}{RMSD} & \multicolumn{1}{c|}{MAE} & \multicolumn{1}{c|}{\makecell{contact\\map\\ accuracy}} & \multicolumn{1}{c|}{RMSD} & \multicolumn{1}{c|}{MAE} \\ \hline
\makecell{Ce\textsubscript{4}Se\textsubscript{8}}   & 0.8                                     & 0.238                     & 0.160                    &    1.0                                       & 0.216                     & 0.149                    &                                  25\%         & 9.24\%                   & 6.88\%                  \\ \hline
\makecell{Tb\textsubscript{2}Cd\textsubscript{4}F\textsubscript{16}} & 0.778                                     & 0.225                     & 0.151                    & 0.909                                         & 0.217                     & 0.124                    &                                  16.84\%         & 3.56\%                   & 17.88\%                  \\ \hline
\makecell{Fe\textsubscript{4}O\textsubscript{4}F\textsubscript{4}}    & 0.667                                     & 0.196                     & 0.127                    & 1.000                                          &  0.124                    & 0.075                    &                                  49.93\%         & 36.73\%                   & 40.94\%                  \\ \hline
\makecell{Nb\textsubscript{4}Cr\textsubscript{4}O\textsubscript{16}}       & 0.692                                     & 0.309                     & 0.183                    & 0.857                                          & 0.190                     & 0.137                    &                                 23.84\%          & 38.51\%                   & 25.14\%                  \\ \hline

\makecell{K\textsubscript{1}Ti\textsubscript{6}Se\textsubscript{8}}       & 0.667                                     & 0.117                     & 0.069                    & 1.000                                          & 0.072                     & 0.051                    &                                 49.93\%          & 38.46\%                   & 26.09\%                  \\ \hline
\makecell{Cr\textsubscript{3}N\textsubscript{6}}   & 0.75                                     & 0.184                     & 0.130                    &    0.8                                       & 0.111                     &  0.089                   &                                 6.67\%          & 39.67\%                   & 31.54\%                  \\ \hline
\makecell{In\textsubscript{2}I\textsubscript{6}O\textsubscript{18}}    & 0.593                                     & 0.191                     & 0.143                    & 0.815                                          & 0.189                     & 0.151                    &                                 37.44\%          & 1.05\%                   & -5.59\%                  \\ \hline
\makecell{Ho\textsubscript{2}H\textsubscript{6}O\textsubscript{6}}      & 0.818                                     &  0.121                    & 0.076                    & 0.9                                          & 0.103                     & 0.065                    &                                 10.02\%          & 14.88\%                   & 14.47\%                  \\ \hline
\makecell{Li\textsubscript{2}Cr\textsubscript{2}O\textsubscript{4}}    & 0.75                                     & 0.265                     & 0.155                    & 1.000                                          & 0.243                     & 0.137                    &                                33.33\%           & 8.3\%                  & 11.61\%                  \\ \hline
\makecell{La\textsubscript{3}Al\textsubscript{1}N\textsubscript{1}} & 1.000                                     & 0.000                     & 0.000                    & 1.000                                          & 0.000                     & 0.000                    &                                0\%             & 0\%                   & 0\%                  \\ \hline
\end{tabular}
\end{center}
\end{table}

\section{Discussion}
\label{sec:others}

Starting with the generated highly symmetric random crystal structures, our algorithm takes the contact map as the optimization target, searches for non-special coordinates at the Wyckoff positions, and achieves successful predictions of a set of high-symmetry crystals based on the contact map with high contact map accuracy, and low RMSD and MAE errors between the predicted Wyckoff positions and the real Wyckoff positions. During the experiment, we found that for crystal structures with extremely high symmetry (e.g. space groups 221, 225), the WPs of the random crystal structure generated by PyXtal contain many special coordinates. While it is easy to obtain high contact map accuracy for these targets, the space for optimization is small and it is difficult to improve the accuracy of the contact map. One solution to address this issue is to generate more random crystal structures, and then screen out suitable random crystal structures with high contact map accuracy. Another notable issue is that in current experiments, we used the real contact map, space group and other information of the target structures in our crystal structure reconstruction. However, in actual situations, the contact map, space group and other information are all predicted for a given composition or material formula, which themselves may contain certain degree of errors, which may affect our reconstruction algorithm's performance. At present, the space group and lattice constants of hypothetical crystal structures can be predicted, and the contact map of a given material formula can also be predicted by the deep learning prediction method\cite{hu2021alphacrystal}. While these methods are still in its emerging stage, they have the potential to achieve high precision crystal structure prediction for any given composition when combined with DFT based fine-tuning.

\section{Conclusion}
\label{sec:others}

In our previous work, we proposed a contact map-based crystal structure prediction framework, which uses global optimization algorithms such as GA and PSO to maximize the match between the predicted contact map of the structure and the contact map of the real crystal structure to search for the coordinates at Wyckoff positions. We showed that geometric constraints such as the contact map or distance matrix of the crystal structure can be utilized for crystal structure reconstruction. However, our CMCrystal algorithm has difficulty to be used for reconstructing the structures of high symmetry crystal materials due to the challenges of obtaining consistent contact map dimension and the difficulty to find exact special coordinates. Here, we propose a new algorithm CMCrystalHS to use PyXtal to generate a random crystal structure with given symmetry constraints based on information such as chemical formulas and space groups. With the contact map as the optimization goal, only the non-special coordinates at the Wyckoff positions are optimized to solve the high symmetry crystal structure prediction problem. We use the contact map accuracy, the root mean square distance (RMSD) and the mean absolute error (MAE) between the predicted Wyckoff positions of the crystal structure and the Wyckoff positions of the target structure to evaluate the reconstructed high symmetry crystal structures. The experimental results show that by using random crystal structures with given symmetry constraints to search for non-special coordinates with the contact map as the optimization target, our proposed method for predicting highly symmetrical crystal structures can achieve high contact map accuracy and the RMSD and MAE between the predicted Wyckoff positions and the real Wyckoff positions are relatively small. This demonstrates our algorithm can be used to solve high-symmetric crystal structure prediction problems based on predicted contact map. When combined with contact map prediction and composition generation algorithms, it will enable large-scale crystal structure prediction and new  materials discovery. 

\section{Availability of data}

The data that support the findings of this study are openly available in Materials Project database at \href{http:\\www.materialsproject.org}{\textcolor{blue}{http:\\www.materialsproject.org}}

\section{Contribution}
Conceptualization, J.H.; methodology, J.H., W.Y.; software, W.Y., J.H ; validation, W.Y.,E.S., J.H.;  investigation, J.H., W.Y., Y.L., E.S., R.D.; resources, J.H.; data curation, J.H., and W.Y.; writing--original draft preparation, Y.Z. and J.H. ; writing--review and editing, J.H, W.Y.; visualization, W.Y and R.D.; supervision, J.H.;  funding acquisition, J.H.

\section{Acknowledgement}
Research reported in this work was supported in part by NSF under grants 1940099 and 1905775. The views, perspective, and content do not necessarily represent the official views of NSF. 

\bibliographystyle{unsrt}  
\bibliography{references}  
\end{document}